\documentclass[pdftex,12pt]{article}
\usepackage{amsmath,amsfonts,amsbsy,latexsym,amssymb,amscd,amstext}
\usepackage{pst-node}
\usepackage{dsfont}
\usepackage{cool}
\usepackage{overpic,youngtab}
\pdfoutput=1
\usepackage{rotating}
\usepackage{tikz}
\usepackage{slashed}

\DeclareMathOperator\arctanh{arctanh}

\def\bpm{\begin{pmatrix}}
\def\epm{\end{pmatrix}}
\def\be{\begin{equation}}

\def\ee{\end{equation}}
\def\bea{\begin{eqnarray}}
\def\eea{\end{eqnarray}}
\def\pd{\partial}
\def\a{\alpha}

\def\b{\beta}

\def\d{\delta}
\def\m{\mu}

\def\n{\nu}
\def\t{\tau}

\def\r{\rho}

\def\s{\sigma}

\def\bma{\begin{pmatrix}}
\def\ema{\end{pmatrix}}

\def\bi{\begin{itemize}}
\def\ei{\end{itemize}}

\begin{document}

		\vspace*{-1cm}
		\phantom{hep-ph/***} 
		{\flushleft
			{{FTUAM-21-7}}
			\hfill{{ IFT-UAM/CSIC-21-96}}}
		\vskip 1.5cm
		\begin{center}
		{\LARGE\bfseries   Unimodular Cosmological models.}\\[3mm]
			\vskip .3cm
		
		\end{center}

		\vskip 0.5  cm
		\begin{center}
			{\large Enrique \'Alvarez and Jes\'us Anero.}
			\\
			\vskip .7cm
			{
				Departamento de F\'isica Te\'orica and Instituto de F\'{\i}sica Te\'orica, 
				IFT-UAM/CSIC,\\
				Universidad Aut\'onoma de Madrid, Cantoblanco, 28049, Madrid, Spain\\
				\vskip .1cm

				\vskip .5cm
				
				\begin{minipage}[l]{.9\textwidth}
					\begin{center} 
							\textit{E-mail:} 
						\tt{enrique.alvarez@uam.es},
						\tt{jesusanero@gmail.com}
					\end{center}
				\end{minipage}
			}
		\end{center}
	\thispagestyle{empty}
	
\begin{abstract}
It is claimed that in the unimodular gravity framework the observational fact of exponential expansion of the universe cannot be taken as evidence for the presence for a cosmological constant or similar quintessence.
\end{abstract}

\newpage
\tableofcontents
	\thispagestyle{empty}
\flushbottom

\newpage

 \section{Introduction. }
 Unimodular gravity (UG) \cite{Einstein} is a theory first proposed by Einstein in 1919 which is identical to general relativity (GR) \textit{except} that we assume the spacetime metric to be unimodular ($det g_{\m\n}\equiv g=-1$). This in turn restricts the symmetry group to transverse diffeomorphisms (those such that their generating vector field obeys $\partial_\m\xi^\m = 0$). Also the equations of motion (EM) are not the full set of Einstein's equations, but only their traceless part (see Appendix \eqref{A}). 
 
 Although it has been sometimes claimed that UG is completely equivalent to
 GR, this is not so. The aim of the present paper is to highlight some crucial differences in a simple cosmological comtext.
 

The Friedmann-Lema\^itre metric  in the unimodular gauge of General Relativity (which is the only admissible metric in Unimodular Gravity  cf. for example \cite{redux})  reads
\be
ds^2=b(t)^{- 3/2}\,dt^2-b(t)^{1/2}\, \d_{ij}dx^i dx^j
\ee
where the function $b$ depends on time only, $b=b(t)$. The cosmic normalized four velocity vector field, $u^\m u_\m=1$, is given explicitly by
\be u^{\m}=\left(b^{3/4},0,0,0\right)\ee
and the projector onto the 3-space is given by
\be
h^\m_\n\equiv \d^\m_\n-u^\m u_\n=\begin{pmatrix}0&0&0&0\\0&1&0&0\\0&0&1&0\\0&0&0&1\end{pmatrix}
\ee
it is easy to check that this  congruence is geodesic
\be
\label{c}u^\n\nabla_\n u^\m=0
\ee
and the volumen expansion reads
\be \label{v1}\theta\equiv \nabla_\m u^\m=\frac{3}{4}b^{-1/4}\frac{db}{dt}\ee

\par
The equation of motion in  UG (the traceless piece of Einstein's) reads
\be 
R_{\m\n}-\dfrac{1}{4}R\,g_{\m\n}=2\kappa^2 \left(T_{\m\n}-\dfrac{1}{4}\,T\,g_{\m\n}\right)\label{UGEM} 
\ee
where the scalar curvature
\be\label{RR} R=-\frac{3}{8\sqrt{b}}\left[\left(\frac{db}{dt}\right)^2+4b\frac{d^2 b}{dt^2}\right]\ee
and we assume matter as a perfect fluid, that is
\be
T_{\m\n}\equiv \left(\r+p\right)u_\m u_\n-p g_{\m\n}
\ee
the energy-momentum conservation, $\nabla_\n T^{\m\n}=0$ is then equivalent to 
\be
u^\m\nabla_\m{\r}+\left(\r+p\right)\theta=0
\ee
using the equation of motion \eqref{UGEM} and the property \eqref{c}, the Raychaudhuri's equation \cite{Raychaudhuri:1955} reduces to
\be u^\m\nabla_\m{\theta}+\frac{1}{n-1}\theta^2+\s_{\a\b}\s^{\a\b}-\omega_{\a\b}\omega^{\a\b}+\frac{1}{n}R+\frac{2(n-1)}{n}\kappa^2(\r+p)=0\ee
with \eqref{RR}, we have
\be R=-2u^\m\nabla_\m{\theta}-\frac{4}{3}\theta^2\ee
assuming as usual for simplicity vanishing shear and rotation, $\s_{\a\b}=\omega_{\a\b}=0$, in the physical dimension $n=4$ 
\be\label{Re} u^\m\nabla_\m{\theta}+3\kappa^2(\r+p)=0\ee
it is worth remarking that it is not possible to express $R$ in terms of $T$.
\par
We can use  the Ellis'clever trick \cite{Ellis:1971pg} is to define a length scale through
\be \label{v2}\theta=\frac{3}{l}u^\m\nabla_\m l\ee
actually
\be
b\sim l^4.
\ee
finally we can write the Raychaudhuri's equation \eqref{Re} like
\be\label{le} u^\m u^\n\Big[ l\nabla_\n\nabla_\m{l}-\nabla_\m l\nabla_\n l\Big]+\kappa^2(\r+p)l^2=0\ee
\subsection{Vacuum solutions.}
Then  $p=\r=0$ and Raychaudhuri's equation reduces to
\be u^\m\nabla_\m{\theta}=0\ee
but this is easy to calculate
\be u^\m\nabla_\m{\theta}=u^\n\nabla_\n\nabla_\m u^\m=-\frac{3}{16\sqrt{b}}\left[\left(\frac{db}{dt}\right)^2-4b\frac{d^2 b}{dt^2}\right]\ee
we obtain the equation of motion for UG
\be\left(\frac{db}{dt}\right)^2-4b\frac{d^2 b}{dt^2}=0\ee

Its general solution is given by either
\be
b=H_0^{4 \over 3}
\ee
(a constant) which corresponds to flat space; or else
\be 
b(t)=H_0^{4 \over 3}\left(3t-t_0\right)^{4\over 3}
\ee
which corresponds to  {\em de Sitter} \footnote{
	In unimodular coordinates, the maximally symmetric, constant curvature {\em de Sitter} spacetime reads
	\be\label{CC}
	ds^2=\left({dt\over 3 H t}\right)^2-(3 H t)^{2/3} \d_{ij}\,dx^i dx^j
	\ee
	that is, precisely
	\be
	b(t)\sim \,t^{4\over 3}
	\ee} space; $H_0\equiv 3\theta$ being the constant expansion. In this solution it is arbitrary, because there is no physical scale in the problem that determines it. It is to be emphasized that this solution depends on two parameters, whereas the  at space solution depends only on one, being thus less generic.
	\par
	This could be anticipated, because the vacuum EM in unimodular gravity are just Einstein spaces 
	\be
	R_{\m\n}={1\over 4} R g_{\m\n}
	\ee
	flat space is just a quite particular solution; constant curvature space-times \cite{Wolf} are a more generic one.
\subsection{Perfect fluid solution.}
Assuming the equation of state
\be p=\omega\r\ee
using the energy-momentum conservation
\be u^\m\Big[\frac{\nabla_\m\r}{\r}+3(1+\omega)\frac{\nabla_\m l}{l}\Big]=0\ee
which solution is
\be \r=\r_0\left(\frac{l}{l_0}\right)^{-3(1+\omega)}\ee
therefore,  we will write the differential equation \eqref{le}
\be\label{lpf}
\frac{1}{l}{d^2 l\over dt^2}-\frac{1}{l^2}\left({dl\over dt}\right)^2+\kappa^2(1+\omega) m_0 l^{-3(3+\omega)}=0
\ee
where $m_0=\r_0l_0^{3(3+\omega)}$. When $\kappa\neq 0$  this equation has a general solution 
\be l(t)=\Big\{\frac{\kappa^2(1+\omega) m_0}{(3+\omega)C_2}\sinh^2\Big[\frac{3\sqrt{C_2}}{2}(3+\omega)(t+C_3)\Big]\Big\}^{\tfrac{1}{3(3+\omega)}}\ee

Now, we can study some special cases
\bi
\item   Let us begin with  $\omega=-1$, corresponding to vacuum energy, in this case, the EM  reduces to the vacuum solution
\be 
b(t)=H_0^{4/3}\,\left(3t-t_0\right)^{4\over 3}
\ee
id est, the generic vacuum solution found earlier.

We can write in function of $l(t)$ like
\be l_{\Lambda}(t)=l_0 e^{H_0t}\label{l1}\ee

The equivalence between the vacuum energy and the vacuum solution is a trivial consecuence of the EM, because the traceless piece of the energy-momentum tensor only depends on the combination $p+\r$ and not on either the pressure or energy density separately. 
This fact is one of the most important physical consequences of unimodularity and in fact is characteristic of it. More on this in the following.

\item Consider now radiation, $\omega=\frac{1}{3}$. The general solution reads
\be l_r(t)=e^{-\sqrt{C_2}(t+C_3)}\Big\{\frac{1}{10C_2}\Big[e^{10\sqrt{C_2}(t+C_3)}-\frac{20}{3}\kappa^2 m_0C_2\Big]\Big\}^{1/5}\ee

\item  When in the pressureless ({\em dust}) fluis, where $\omega=0$, the general solution is easily found to be
\be l_d(t)=e^{-\sqrt{C_2}(t+C_3)}\Big\{\frac{1}{6C_2}\Big[e^{9\sqrt{C_2}(t+C_3)}-2\kappa^2 m_0C_2\Big]\Big\}^{2/9}\ee

\item Finally $\omega=-\frac{1}{3}$ (sometimes called {\em quintessence}) the full solution reads
\be l_q(t)=e^{-\sqrt{C_2}(t+C_3)}\Big\{\frac{1}{4C_2}\Big[e^{8\sqrt{C_2}(t+C_3)}-\frac{2}{3}\kappa^2 m_0C_2\Big]\Big\}^{1/4}\ee
\ei
\subsection{The line $p+\r=C$.}

Let us examine the inhomogeneous equation of state $\r+p=C$. This is one of the most interesting results of UG, where physics depends on the value of the constant $C$ only.
\par
Indeed, depending on the value of the constant $C$, it is possible that both $p$ and $\r$ are positive. Only in the case $C=0$ is this situation strictly equivalent to a vacuum energy density. We will write the differential equation \eqref{le}
\be
\frac{1}{l}{d^2 l\over dt^2}-\frac{1}{l^2}\left({dl\over dt}\right)^2+C\kappa^2l^{-4}=0
\ee
and the general solution reads
\be l_C(t)=e^{-\sqrt{C_2}(t+C_3)}\Big\{\frac{1}{6C_2}\Big[e^{6\sqrt{C_2}(t+C_3)}-3C\kappa^2C_2\Big]\Big\}^{1/3}\ee
obviously when $C=0$ this solution reduces to the vacuum solution
\be l_{\Lambda}(t)=l_0 e^{H_0t}\ee
with $C_3=0$ and $C_2^2=H_0$. 
\par
More interestingly, this solution is an atractor asymptotically when $t\rightarrow\infty$. Any solution tends asymptotically to de Sitter.
\par
For $C C_2 >0$ there is an {\em origin of time} $t_0$. For earlier times $t< t_0$ the solution becames unphysical. To be specific,
\be
t_0={1\over 6 \sqrt{C_2}}\log\,\left(3 C \kappa^2 C_2\right)- C_3
\ee
\section{Unimodular gravity versus General Relativity.}
The unimodular gauge of General Relativity (GR) is of course fully equivalent to the usual formulation of GR in comoving coordinates \cite{Weinberg, Bondi}  where the metric reads
\be\label{mc}
ds^2= d\t^2-a(\t)^2 \sum\d_{ij} dx^i dx^j
\ee
with a four velocity
\be u^\m=(1,0,0,0)\ee
and 
\be u^\n\nabla_\n{u^\m}=0\ee
in this case 
\be 
\theta=3\frac{1}{a}\frac{da}{dt}
\ee

We insist that he only difference between GR and UG stems from the EM. Let is spell this out in some detail.

Now the equation of motion is the usual Einstein one
\be 
R_{\m\n}-\dfrac{1}{2}R\,g_{\m\n}=2\kappa^2 T_{\m\n}\label{EM} 
\ee
in this case the scalar of curvature reads
\be R=-\frac{6}{a^2}\left[\left(\frac{da}{dt}\right)^2+a\frac{d^2a}{dt^2}\right]\ee
then the Raychaudhuri's equation  in comoving coordinates yields
\be -3\left(\frac{da}{dt}\right)^2+2a^2\kappa^2\r=0\ee

In this case, the vaccum solution reduces to
\be \frac{da}{dt}=0\ee
i.e $\theta=0$ which is just flat spacetime, it is a subset of the UG result, $\dot{\theta}=0$
\subsection{Perfect fluid solution.}
Assuming the equation of state
\be 
p=\omega\r
\ee
we easily get
\be 
\r=\r_0\left(\frac{a}{a_0}\right)^{-3(1+\omega)}
\ee
then
\be 
-3\left(\frac{da}{dt}\right)^2+2\kappa^2m_0 a^{-1-3\omega}=0
\ee
where $m_0=\r_0a_0^{3(1+\omega)}$. 

We can study this equation compared with the unimodular one \eqref{lpf}, if we derive
\be -6\frac{da}{dt}\frac{d^2a}{dt^2}-2(1+3\omega)\kappa^2m_0 a^{-2-3\omega}\frac{da}{dt}=0\ee
we omit the trivial case when $a=0$, then we can write.
\be a\frac{d^2a}{dt^2}=-\frac{1}{3}(1+3\omega)\kappa^2m_0a^{-1-3\omega}\ee
finally
\be a\frac{d^2a}{dt^2}-\left(\frac{da}{dt}\right)^2+(1+\omega)\kappa^2m_0a^{-1-3\omega}=0\ee
using \eqref{v2}, we arrive to  \eqref{lpf}, we conclude the comoving solution is  a subset of the unimodular solution.

The general solution of the above reads
\be 
a(t)=\Big\{\frac{1+\omega}{2}\Big[\sqrt{6m_0}\kappa t+3C_2\Big]\Big\}^{\frac{2}{3(1+\omega)}}
\ee

Let us agein specify some special cases in parallel with the ones considered above in UG,
\bi
\item When  $\omega=-1$,
\be 
a_{\Lambda}(t)=a_0 e^{\sqrt{\frac{2\r_0}{3}}\kappa t}
\ee
which coincides with the UG solution, \eqref{l1}, when $H_0=\kappa\sqrt{\frac{2\r_0}{3}}$.

\item For radiation  $\omega=\frac{1}{3}$ we get
\be 
a_r(t)=\sqrt{2\kappa t\sqrt{\frac{\r_0}{3}}a_0^2+C_2}
\ee
\item For presureless matter $\omega=0$ the result is
\be 
a_d(t)=\left(\kappa t\frac{a_0}{2}\sqrt{6a_0\r_0}+C_2\right)^{2/3}
\ee
\item And for  quintessence $\omega=-\frac{1}{3}$ we get
\be 
a_q(t)=t\kappa\sqrt{\frac{2\r_0}{3}}a_0+C_2
\ee

\ei

\section{Scalar field.}
Consider a scalar field minimally coupled to the gravitational field. In unimodular coordinates
\bea
&&S=\int d^4 x \sqrt{|g|}\,\left\{{1\over 2} g^{\m\n} \pd_\m \phi \pd_\n  \phi-V(\phi)\right\}=\nonumber\\
&&=\int d^4 x \left\{{1\over 2}\left(b^{3/2} \dot{\phi}^2-b^{- 1/2} (\pd\phi)^2\right)-V(\phi)\right\}
\eea
where $\dot{\phi}\equiv {\pd\phi\over \pd t}$. The EM for the scalar field reads,
\be 
\Box \phi +V^{\prime}(\phi)=0\label{scalar}
\ee
in other words
\be\label{EMS}
\pd_0\left(b^{3/2}\dot{\phi}\right)-b^{- 1/2}\sum \pd^2_i \phi+V^\prime(\phi)=0
\ee
a solution that respects the isometries of the space cannot depend on the space coordinates, i.e.
\be
\pd_ i \phi=0
\ee
and the canonical energy-momentum tensor  is given by
\be
T_{\m\n}^{can}\equiv \dfrac{\partial \mathcal{L}}{\partial (\pd^\m\phi)}\pd_\n \phi-\mathcal{L} g_{\m\n}.
\ee

On the other hand \cite{Erickson} for scalar lagrangians of the quite general form $L(X,\phi)$ where
\be
X\equiv {1\over 2}(\pd_\m\phi)^2
\ee
the velocity of sound is
\be
c_s^2={{\pd p\over \pd X}\over {\pd \r\over \pd X}}
\ee
It so happens that
\be
w={X-V\over X+V}
\ee
and 
\be
c_s^2=1
\ee
\par
The UG equation of motion corresponding to this source read
\bea\label{EMUG}
&&-{3\over 8}{\ddot{b}\over b}+{3\over 32}{\dot{b}^2\over b^2}={3\over 2}\kappa^2 \dot{\phi}^2\nonumber\\
&&\left({\dot{b}^2\over 32}-{b \ddot{b}\over 8}\right)\d_{ij}=\kappa^2 {b^2\over 2}\dot{\phi}^2 \d_{ij}.
\eea
again $\dot{b}=\frac{db}{dt}$. Let us find a general solution for a free field. The EM of the scalar field \eqref{EMS}, implies
\be b^{3/2}\dot{\phi}=C\label{CTEM}\ee
solving  for $\ddot{b}$ the second equation \eqref{EMUG}, we get
\be
\dfrac{b\dot{b}^2}{4}-b^2\ddot{b}=4\kappa^2C^2
\ee
which is an identity for the first equation. The general solution reads

\be 
b(t)=\frac{1}{4}\Bigg[\frac{-256\kappa^2C^2+27C_1^2(t+C_2)^2}{6C_1}\Bigg]^{2/3}
\ee
using \eqref{CTEM} we obtain the correspondent scalar field
\be
\phi(t)=-\frac{1}{\kappa\sqrt{3}}\arctanh\Bigg[\frac{3\sqrt{3}C_1(t+C_2)}{16\kappa C}\Bigg]+C_3.\
\ee

Consider the case when the initial conditions at time $t=0$ are such that the unimodular scale factor vanishes 
\bea\label{BC}
&b(0)=0
\eea
then the solution reduces to
\be
b(t)={1\over 4}\,\left({9C_1\over 2}\right)^{2\over 3}\,t^{4\over 3}
\ee
the constant scalar spacetime curvature corresponding to \eqref{CC} is negative 
with our conventions
\be
R=-12 H_0^2.
\ee
\par
To summarize, we have shown that with boundary conditions as above \eqref{BC} the solution for gravity sourced by a free scalar field corresponds to an
exponential expansion with Hubble parameter 
\be
H=\frac{\sqrt{C_1}}{4}={1\over 4\, b^{1/4}}\,{d b\over dt}
\ee

\section{Conclusions.}
The fact that the equations of motion corresponding to Einstein's  unimodular gravity  are just the traceless piece of Einstein's usual General Relativity ones  means that there is a degeneracy insofar as the source term  only depends on the combination $p+\r$ instead of depending on the pressure and energy density separately.\\
The main consequence of the above is that as far as UG is concerned there is no distinction between vacuum and cosmological constant. The vacuum cosmological solution is  determined not by the initial condition $b(t_0)$ as in GR, but by the couple $\left(b(t_0),\dot{b}(t_0)\right)$. What makes the vacuum universe exponentially expand is simply the initial condition $\dot{b}(t_0)\neq 0$.
\par
This means for example that the observational fact of exponential expansion of the universe cannot be taken as evidence  for the presence for a cosmological constant or similar quintessence.
\par
In fact the physical difference between both theories is more general. 
\par
In particular, UG yields the same gravitational field for all perfect fluids in the line $p+\r=C$. This is {\em not} a particular equation of state insofar as it is an inhomogeneour relation between the fluid variables. 
\par
Please note that for any strictly positive constant $C>0$ both energy and pressure can be positive (although they do not need to do so in order to stay in the line). Nevertheless exponential expansion is an atractor in the sense that all solutions in this line converge asymptotically to it when the time goes to infinity. 

This degeneracy is also the explanation of one aspect of the cosmological constant problem \cite{windows}, namely why the supposedly huge QFT vacuum energy  does not imply an equally  huge cosmological constant.

Indeed, from the UG viewpoint $p+\rho=0$ (vacuum energy) is just indistinguishable  to plain vacuum. May be this would lead to a way to tell apart Einstein's two gravitational EM, namely general relativity versus unimodular gravity.

\section{Acknowledgements.}
One of us (EA) acknowledges the warm hospitality of Robert Brandenberger, Carlos Arg\"uelles and Tracy Slatyer at MacGill, Harvard and MIT while this work was begun. We acknowledge partial financial support by the Spanish MINECO through the Centro de excelencia Severo Ochoa Program under grant SEV-2016-0597, by the Spanish ``Agencia Estatal de Investigac\'ion''(AEI) and the EU ``Fondo Europeo de Desarrollo Regional'' (FEDER) through the project PID2019-108892RB-I00/AEI/10.13039/501100011033. All authors acknowledge the European Union's Horizon 2020 research and innovation programme under the Marie Sklodowska-Curie grant agreement No 860881-HIDDeN. VS acknowledges support from the UK STFC via Grant ST/L000504/1.

\newpage
\appendix
\section{UG versus GR($\lambda$).}\label{A}

We have already mentioned that there is a difference between UG and GR, even in the unimodular gauge for the latter, namely, that the UG equations of motion  include only the {\em traceless} part of Einstein's equations. To be specific, the equation of motion for UG \cite{AGMM} reads
\be\label{UG} 
R_{\m\n}-\dfrac{1}{4}R\,g_{\m\n}=2\kappa^2 \left(T_{\m\n}-\dfrac{1}{4}\,T\,g_{\m\n}\right). 
\ee
\par
It is worth highlighting the fact   that the equations of motion for UG \eqref{UG} are enough by themselves to determine the metric. It has been pointed out that in order for this equation to be compatible with the contracted form of Bianchi's identity.
\be\label{FI}
\nabla^\m R_{\m\n}={1\over 2}\nabla_\n R
\ee
it is necessary that
\be\label{C}
\nabla_\m\left(R+2\kappa^2 T\right)=0.
\ee
which is equivalent to
\be\label{trace}
R+2\kappa^2 T=C= 4 \Lambda
\ee
with our conventions. The rationale for equating the second member with (four times) the cosmological constant stems from the fact that in  GR the trace of Einstein's equations yields 
\be
{n-2\over 2} R+2\kappa^2 T= n \Lambda
\ee
Bianchi's identities, as implicit by their name are identically fulfilled by any spacetime metric, and any energy-momentum tensor derived from a diffeomorphism invariant action principle is also covariantly conserved as well. 
\par
What happens is that not all components of Einstein's equations are linearly independent; there are nontrivial linear combinations of them  that are zero. Bianchi's identities precisely do the spelling as to what are those linear combinations in detail. 
\par
This is related to the fact \cite{Bij} that when defining the flat space spin 2 massless theory starting from the massive one, we are only in need of three gauge symmetries in order to reduce the number of degrees of freedom from 5 (corresponding to a massive spin 2 particle) down to 2 (corresponding to a massless particle). This suggests that 4 gauge symmetries, as in the Fierz-Pauli theory, namely the four independent components of a vector field $\xi^\m$, is an overkill, and that is enough with the three independent components of a transverse vector field, $\pd_\m \xi^\m_T=0$, precisely the generator of the UG symmetry group. Indeed the traceless part of Einstein's equations, are all linearly independent.
\newpage

\end{document}